\documentclass[12pt]{article}
\usepackage[english]{babel}
\usepackage{epsfig,a4}
\begin{document}
%
%
%
\newcommand{\Zo}{Z}
\newcommand{\Zuds}{Z~$\rightarrow$~u$\overline{{\rm u}},$d$\overline{{\rm d}},$s$\overline{{\rm s}}$}
\newcommand{\Zbb}{Z~$\rightarrow$~b$\overline{{\rm b}}$ }
\newcommand{\Zcc}{Z~$\rightarrow$~c$\overline{{\rm c}}$ }
\newcommand{\gcc}{g~$\rightarrow$~c$\overline{{\rm c}}$ }
\newcommand{\ms}   {$\;$}
\newcommand{\Kol}  {K$^0_{\rm L}$}
\newcommand{\Kos}  {K$^0_{\rm S}$}
\newcommand{\Ko}   {K$^0$}
\newcommand{\Kobar}  {${\rm \overline{K}}^0$}
\newcommand{\Kpm}  {K$^\pm$}
\newcommand{\Kp}   {K$^+$}
\newcommand{\Km}   {K$^-$}
\newcommand{\pipm} {$\pi^\pm$}
\newcommand{\pip}  {$\pi^+$}
\newcommand{\pim}  {$\pi^-$}
\newcommand{\pio}  {$\pi^0$}
\newcommand{\p}    {$p$}
\newcommand{\pbar} {$\overline{p}$}
\newcommand{\mupm} {$\mu^\pm$}
\newcommand{\epm}  {$e^\pm$}
\newcommand{\Do}     {D$^0$}
\newcommand{\Dp}     {D$^+$}
\newcommand{\ds}     {D$_{\rm s}$}
\newcommand{\dsp}    {D$_{\rm s}^+$}
\newcommand{\dsm}    {D$_{\rm s}^-$}
\newcommand{\dspm}   {D$_{\rm s}^\pm$}
\newcommand{\dsst}  {D$_{\rm s}^{\ast}$}
\newcommand{\dsstp}  {D$_{\rm s}^{\ast +}$}
\newcommand{\dstpm}  {D$^{\ast\pm}$}
\newcommand{\dst}   {D$^{\ast}$}
\newcommand{\dstst}   {D$^{\ast\ast}$}
\newcommand{\dstp}   {D$^{\ast+}$}
\newcommand{\dsto}   {D$^{\ast 0}$}
\newcommand{\dsd}      {D$_{\rm s}^{\ast\ast}$}
\newcommand{\dsdp}     {D$_{\rm s}^{\ast\ast+}$}
\newcommand{\dsdm}     {D$_{\rm s}^{\ast\ast-}$}
\newcommand{\dsdpm}    {D$_{\rm s}^{\ast\ast\pm}$}
\newcommand{\dsone}    {D$_{\rm s1}$}
\newcommand{\dsonep}   {D$_{\rm s1}^+$}
\newcommand{\dsonem}   {D$_{\rm s1}^-$}
\newcommand{\dsonepm}  {D$_{\rm s1}^\pm$}
\newcommand{\dsonest}   {D$_{\rm s1}^{\ast}$}
\newcommand{\dsonestp}  {D$_{\rm s1}^{\ast+}$}
\newcommand{\dsonestm}  {D$_{\rm s1}^{\ast-}$}
\newcommand{\dsonestpm} {D$_{\rm s1}^{\ast\pm}$}
\newcommand{\dszerost}  {D$_{\rm s0}^{\ast}$}
\newcommand{\dszerostp} {D$_{\rm s0}^{\ast+}$}
\newcommand{\dszerostm} {D$_{\rm s0}^{\ast-}$}
\newcommand{\dszerostpm}{D$_{\rm s0}^{\ast\pm}$}
\newcommand{\dstwost}   {D$_{\rm s2}^{\ast}$}
\newcommand{\dstwostp}  {D$_{\rm s2}^{\ast+}$}
\newcommand{\dstwostm}  {D$_{\rm s2}^{\ast-}$}
\newcommand{\dstwostpm} {D$_{\rm s2}^{\ast\pm}$}
\newcommand{\arw}{$\rightarrow$}
\newcommand{\docha}  {\Do~\arw~\Km \pip}
\newcommand{\dochbo}  {\Do~\arw~\Kobar \pip \pim}
\newcommand{\dochb}  {\Do~\arw~\Kobar \pip \pim}
\newcommand{\dochc}  {\Do~\arw~\Km \pip \pim \pip}
\newcommand{\dochd}  {\Do~\arw~\Km \pip \pio}
\newcommand{\dsda}  {\dsonep~\arw~\dstp \Ko}
\newcommand{\dsdb}  {\dsonep~\arw~\dsto \Kp}
\newcommand{\dsdc}  {\dstwostp~\arw~\Do \Kp}
\newcommand{\dmone}   {$\Delta$M$_1$ = m(\Do,\pip)$-$m(\Do)}
\newcommand{\dmtwo}   {$\Delta$M$_2$ = 
                         m(\dstp,\Ko)$-$m(\dstp)$-$m(\Ko)+m$_{nom}$(\Ko)}
\newcommand{\dmthree} {$\Delta$M$_3$ = m(\Do,\Kp)$-$m(\Do)}
\newcommand{\Mev}     {\mbox{{\rm MeV}}}
\newcommand{\Mevc}    {\mbox{{\rm MeV}/$c$}$\,$}
\newcommand{\Mevcc}   {\mbox{{\rm MeV}/$c^2$}$\,$}
\newcommand{\Mevccns} {\mbox{{\rm MeV}/$c^2$}}
\newcommand{\Gev}     {\mbox{\rm GeV}$\,$}
\newcommand{\Gevc}    {\mbox{{\rm GeV}/$c$}$\,$}
\newcommand{\Gevcc}   {\mbox{{\rm GeV}/$c^2$}$\,$}
\newcommand{\Br}   {\mbox{\rm Br}}
\newcommand{\bigtimes}   {\mbox{\large $\times$}}

\newcommand{\Pevt}      {${\cal P}_{\mathrm{evt}}$}

\centerline{\large EUROPEAN ORGANIZATION FOR NUCLEAR RESEARCH}
\vspace{15mm}
\begin{flushright}
CERN--EP/2001--085\\
28th November 2001
\end{flushright}

\noindent
\vspace{30mm} \\
\centerline{\LARGE\bf\boldmath Production of \mbox{{\dsd} mesons}} 
\vspace{2mm} \\
\centerline{\LARGE\bf in hadronic \Zo~decays} 
\vspace{4mm} \\
\begin{center}
{\em The ALEPH Collaboration
\footnote{See the following pages for the list of authors.}
}\\
\vspace{12mm}
\begin{abstract} \noindent
The production rates of the orbitally excited \dsd~mesons, 
{\dsonepm} and {\dstwostpm}, are measured with the    
 4.1 million hadronic \Zo~decays recorded by the 
ALEPH detector during 1991--1995.
The \dsd~mesons are reconstructed in the decay modes 
{\dsda}, {\dsdb} and {\dsdc}.
The production rate of the {\dsonepm} is measured to be 
\mbox{{\it f}$\,$(\Zo ~\arw~ \dsonepm) $=$} \mbox{$(0.52 \pm 0.09 \pm 0.06)\%$},
under the assumption that the two considered 
decay modes of the {\dsonepm} saturate the branching ratio.
The production rate of the {\dstwostpm} is determined to be
\mbox{{\it f}$\,$(\Zo ~\arw~ \dstwostpm)
 $=(0.83 \pm 0.29 {^{+0.07}_{-0.13}})\%$}, assuming that
the branching fraction of the decay {\dsdc} is 45\%. The production rates 
in \Zcc~and \Zbb~decays are measured separately.
\end{abstract}
 
\vfill
   {(submitted to Phys. Lett. )}
\vspace{15mm}
\end{center}

\newpage \pagestyle{plain} \setcounter{page}{1}

\pagestyle{empty}
\newpage
\small
%
\newlength{\saveparskip}
\newlength{\savetextheight}
\newlength{\savetopmargin}
\newlength{\savetextwidth}
\newlength{\saveoddsidemargin}
\newlength{\savetopsep}
\setlength{\saveparskip}{\parskip}
\setlength{\savetextheight}{\textheight}
\setlength{\savetopmargin}{\topmargin}
\setlength{\savetextwidth}{\textwidth}
\setlength{\saveoddsidemargin}{\oddsidemargin}
\setlength{\savetopsep}{\topsep}
%
%
\setlength{\parskip}{0.0cm}
\setlength{\textheight}{25.0cm}
\setlength{\topmargin}{-1.5cm}
\setlength{\textwidth}{16 cm}
\setlength{\oddsidemargin}{-0.0cm}
\setlength{\topsep}{1mm}
\pretolerance=10000
\centerline{\large\bf The ALEPH Collaboration}
\footnotesize
\vspace{0.5cm}
{\raggedbottom
\begin{sloppypar}
\samepage\noindent
A.~Heister,
S.~Schael
\nopagebreak
\begin{center}
\parbox{15.5cm}{\sl\samepage
Physikalisches Institut das RWTH-Aachen, D-52056 Aachen, Germany}
\end{center}\end{sloppypar}
\vspace{2mm}
\begin{sloppypar}
\noindent
R.~Barate,
I.~De~Bonis,
D.~Decamp,
C.~Goy,
\mbox{J.-P.~Lees},
E.~Merle,
\mbox{M.-N.~Minard},
B.~Pietrzyk
\nopagebreak
\begin{center}
\parbox{15.5cm}{\sl\samepage
Laboratoire de Physique des Particules (LAPP), IN$^{2}$P$^{3}$-CNRS,
F-74019 Annecy-le-Vieux Cedex, France}
\end{center}\end{sloppypar}
\vspace{2mm}
\begin{sloppypar}
\noindent
G.~Boix,
S.~Bravo,
M.P.~Casado,
M.~Chmeissani,
J.M.~Crespo,
E.~Fernandez,
\mbox{M.~Fernandez-Bosman},
Ll.~Garrido,$^{15}$
E.~Graug\'{e}s,
M.~Martinez,
G.~Merino,
R.~Miquel,$^{27}$
Ll.M.~Mir,$^{27}$
A.~Pacheco,
H.~Ruiz
\nopagebreak
\begin{center}
\parbox{15.5cm}{\sl\samepage
Institut de F\'{i}sica d'Altes Energies, Universitat Aut\`{o}noma
de Barcelona, E-08193 Bellaterra (Barcelona), Spain$^{7}$}
\end{center}\end{sloppypar}
\vspace{2mm}
\begin{sloppypar}
\noindent
A.~Colaleo,
D.~Creanza,
M.~de~Palma,
G.~Iaselli,
G.~Maggi,
M.~Maggi,
S.~Nuzzo,
A.~Ranieri,
G.~Raso,$^{23}$
F.~Ruggieri,
G.~Selvaggi,
L.~Silvestris,
P.~Tempesta,
A.~Tricomi,$^{3}$
G.~Zito
\nopagebreak
\begin{center}
\parbox{15.5cm}{\sl\samepage
Dipartimento di Fisica, INFN Sezione di Bari, I-70126
Bari, Italy}
\end{center}\end{sloppypar}
\vspace{2mm}
\begin{sloppypar}
\noindent
X.~Huang,
J.~Lin,
Q. Ouyang,
T.~Wang,
Y.~Xie,
R.~Xu,
S.~Xue,
J.~Zhang,
L.~Zhang,
W.~Zhao
\nopagebreak
\begin{center}
\parbox{15.5cm}{\sl\samepage
Institute of High Energy Physics, Academia Sinica, Beijing, The People's
Republic of China$^{8}$}
\end{center}\end{sloppypar}
\vspace{2mm}
\begin{sloppypar}
\noindent
D.~Abbaneo,
P.~Azzurri,
O.~Buchm\"uller,$^{25}$
M.~Cattaneo,
F.~Cerutti,
B.~Clerbaux,
H.~Drevermann,
R.W.~Forty,
M.~Frank,
F.~Gianotti,
T.C.~Greening,$^{29}$
J.B.~Hansen,
J.~Harvey,
D.E.~Hutchcroft,
P.~Janot,
B.~Jost,
M.~Kado,$^{27}$
P.~Mato,
A.~Moutoussi,
F.~Ranjard,
L.~Rolandi,
D.~Schlatter,
O.~Schneider,$^{2}$
G.~Sguazzoni,
W.~Tejessy,
F.~Teubert,
A.~Valassi,
I.~Videau,
J.~Ward
\nopagebreak
\begin{center}
\parbox{15.5cm}{\sl\samepage
European Laboratory for Particle Physics (CERN), CH-1211 Geneva 23,
Switzerland}
\end{center}\end{sloppypar}
\vspace{2mm}
\begin{sloppypar}
\noindent
F.~Badaud,
A.~Falvard,$^{22}$
P.~Gay,
P.~Henrard,
J.~Jousset,
B.~Michel,
S.~Monteil,
\mbox{J-C.~Montret},
D.~Pallin,
P.~Perret
\nopagebreak
\begin{center}
\parbox{15.5cm}{\sl\samepage
Laboratoire de Physique Corpusculaire, Universit\'e Blaise Pascal,
IN$^{2}$P$^{3}$-CNRS, Clermont-Ferrand, F-63177 Aubi\`{e}re, France}
\end{center}\end{sloppypar}
\vspace{2mm}
\begin{sloppypar}
\noindent
J.D.~Hansen,
J.R.~Hansen,
P.H.~Hansen,
B.S.~Nilsson,
A.~W\"a\"an\"anen
\begin{center}
\parbox{15.5cm}{\sl\samepage
Niels Bohr Institute, DK-2100 Copenhagen, Denmark$^{9}$}
\end{center}\end{sloppypar}
\vspace{2mm}
\begin{sloppypar}
\noindent
A.~Kyriakis,
C.~Markou,
E.~Simopoulou,
A.~Vayaki,
K.~Zachariadou
\nopagebreak
\begin{center}
\parbox{15.5cm}{\sl\samepage
Nuclear Research Center Demokritos (NRCD), GR-15310 Attiki, Greece}
\end{center}\end{sloppypar}
\vspace{2mm}
\begin{sloppypar}
\noindent
A.~Blondel,$^{12}$
G.~Bonneaud,
\mbox{J.-C.~Brient},
A.~Roug\'{e},
M.~Rumpf,
M.~Swynghedauw,
M.~Verderi,
\linebreak
H.~Videau
\nopagebreak
\begin{center}
\parbox{15.5cm}{\sl\samepage
Laboratoire de Physique Nucl\'eaire et des Hautes Energies, Ecole
Polytechnique, IN$^{2}$P$^{3}$-CNRS, \mbox{F-91128} Palaiseau Cedex, France}
\end{center}\end{sloppypar}
\vspace{2mm}
\begin{sloppypar}
\noindent
V.~Ciulli,
E.~Focardi,
G.~Parrini
\nopagebreak
\begin{center}
\parbox{15.5cm}{\sl\samepage
Dipartimento di Fisica, Universit\`a di Firenze, INFN Sezione di Firenze,
I-50125 Firenze, Italy}
\end{center}\end{sloppypar}
\vspace{2mm}
\begin{sloppypar}
\noindent
A.~Antonelli,
M.~Antonelli,
G.~Bencivenni,
G.~Bologna,$^{4}$
F.~Bossi,
P.~Campana,
G.~Capon,
V.~Chiarella,
P.~Laurelli,
G.~Mannocchi,$^{5}$
F.~Murtas,
G.P.~Murtas,
L.~Passalacqua,
\mbox{M.~Pepe-Altarelli},$^{24}$
P.~Spagnolo
\nopagebreak
\begin{center}
\parbox{15.5cm}{\sl\samepage
Laboratori Nazionali dell'INFN (LNF-INFN), I-00044 Frascati, Italy}
\end{center}\end{sloppypar}
\vspace{2mm}
\begin{sloppypar}
\noindent
A.~Halley,
J.G.~Lynch,
P.~Negus,
V.~O'Shea,
C.~Raine,$^{4}$
A.S.~Thompson
\nopagebreak
\begin{center}
\parbox{15.5cm}{\sl\samepage
Department of Physics and Astronomy, University of Glasgow, Glasgow G12
8QQ,United Kingdom$^{10}$}
\end{center}\end{sloppypar}
\vspace{2mm}
\begin{sloppypar}
\noindent
S.~Wasserbaech
\nopagebreak
\begin{center}
\parbox{15.5cm}{\sl\samepage
Department of Physics, Haverford College, Haverford, PA 19041-1392, U.S.A.}
\end{center}\end{sloppypar}
\vspace{2mm}
\begin{sloppypar}
\noindent
R.~Cavanaugh,
S.~Dhamotharan,
C.~Geweniger,
P.~Hanke,
G.~Hansper,
V.~Hepp,
E.E.~Kluge,
A.~Putzer,
J.~Sommer,
K.~Tittel,
S.~Werner,$^{19}$
M.~Wunsch$^{19}$
\nopagebreak
\begin{center}
\parbox{15.5cm}{\sl\samepage
Kirchhoff-Institut f\"ur Physik, Universit\"at Heidelberg, D-69120
Heidelberg, Germany$^{16}$}
\end{center}\end{sloppypar}
\vspace{2mm}
\begin{sloppypar}
\noindent
R.~Beuselinck,
D.M.~Binnie,
W.~Cameron,
P.J.~Dornan,
M.~Girone,$^{1}$
N.~Marinelli,
J.K.~Sedgbeer,
J.C.~Thompson$^{14}$
\nopagebreak
\begin{center}
\parbox{15.5cm}{\sl\samepage
Department of Physics, Imperial College, London SW7 2BZ,
United Kingdom$^{10}$}
\end{center}\end{sloppypar}
\vspace{2mm}
\begin{sloppypar}
\noindent
V.M.~Ghete,
P.~Girtler,
E.~Kneringer,
D.~Kuhn,
G.~Rudolph
\nopagebreak
\begin{center}
\parbox{15.5cm}{\sl\samepage
Institut f\"ur Experimentalphysik, Universit\"at Innsbruck, A-6020
Innsbruck, Austria$^{18}$}
\end{center}\end{sloppypar}
\vspace{2mm}
\begin{sloppypar}
\noindent
E.~Bouhova-Thacker,
C.K.~Bowdery,
A.J.~Finch,
F.~Foster,
G.~Hughes,
R.W.L.~Jones,
M.R.~Pearson,
N.A.~Robertson
\nopagebreak
\begin{center}
\parbox{15.5cm}{\sl\samepage
Department of Physics, University of Lancaster, Lancaster LA1 4YB,
United Kingdom$^{10}$}
\end{center}\end{sloppypar}
\vspace{2mm}
\begin{sloppypar}
\noindent
K.~Jakobs,
K.~Kleinknecht,
G.~Quast,$^{6}$
B.~Renk,
\mbox{H.-G.~Sander},
H.~Wachsmuth,
C.~Zeitnitz
\nopagebreak
\begin{center}
\parbox{15.5cm}{\sl\samepage
Institut f\"ur Physik, Universit\"at Mainz, D-55099 Mainz, Germany$^{16}$}
\end{center}\end{sloppypar}
\vspace{2mm}
\begin{sloppypar}
\noindent
A.~Bonissent,
J.~Carr,
P.~Coyle,
O.~Leroy,
P.~Payre,
D.~Rousseau,
M.~Talby
\nopagebreak
\begin{center}
\parbox{15.5cm}{\sl\samepage
Centre de Physique des Particules, Universit\'e de la M\'editerran\'ee,
IN$^{2}$P$^{3}$-CNRS, F-13288 Marseille, France}
\end{center}\end{sloppypar}
\vspace{2mm}
\begin{sloppypar}
\noindent
F.~Ragusa
\nopagebreak
\begin{center}
\parbox{15.5cm}{\sl\samepage
Dipartimento di Fisica, Universit\`a di Milano e INFN Sezione di Milano,
I-20133 Milano, Italy}
\end{center}\end{sloppypar}
\vspace{2mm}
\begin{sloppypar}
\noindent
A.~David,
H.~Dietl,
G.~Ganis,$^{26}$
K.~H\"uttmann,
G.~L\"utjens,
C.~Mannert,
W.~M\"anner,
\mbox{H.-G.~Moser},
R.~Settles,
H.~Stenzel,
W.~Wiedenmann,
G.~Wolf
\nopagebreak
\begin{center}
\parbox{15.5cm}{\sl\samepage
Max-Planck-Institut f\"ur Physik, Werner-Heisenberg-Institut,
D-80805 M\"unchen, Germany\footnotemark[16]}
\end{center}\end{sloppypar}
\vspace{2mm}
\begin{sloppypar}
\noindent
J.~Boucrot,
O.~Callot,
M.~Davier,
L.~Duflot,
\mbox{J.-F.~Grivaz},
Ph.~Heusse,
A.~Jacholkowska,$^{22}$
J.~Lefran\c{c}ois,
\mbox{J.-J.~Veillet},
C.~Yuan
\nopagebreak
\begin{center}
\parbox{15.5cm}{\sl\samepage
Laboratoire de l'Acc\'el\'erateur Lin\'eaire, Universit\'e de Paris-Sud,
IN$^{2}$P$^{3}$-CNRS, F-91898 Orsay Cedex, France}
\end{center}\end{sloppypar}
\vspace{2mm}
\begin{sloppypar}
\noindent
G.~Bagliesi,
T.~Boccali,
L.~Fo\`{a},
A.~Giammanco,
A.~Giassi,
F.~Ligabue,
A.~Messineo,
F.~Palla,
G.~Sanguinetti,
A.~Sciab\`a,
R.~Tenchini,$^{1}$
A.~Venturi,$^{1}$
P.G.~Verdini
\samepage
\begin{center}
\parbox{15.5cm}{\sl\samepage
Dipartimento di Fisica dell'Universit\`a, INFN Sezione di Pisa,
e Scuola Normale Superiore, I-56010 Pisa, Italy}
\end{center}\end{sloppypar}
\vspace{2mm}
\begin{sloppypar}
\noindent
G.A.~Blair,
G.~Cowan,
M.G.~Green,
T.~Medcalf,
A.~Misiejuk,
J.A.~Strong,
\mbox{P.~Teixeira-Dias},
\mbox{J.H.~von~Wimmersperg-Toeller}
\nopagebreak
\begin{center}
\parbox{15.5cm}{\sl\samepage
Department of Physics, Royal Holloway \& Bedford New College,
University of London, Egham, Surrey TW20 OEX, United Kingdom$^{10}$}
\end{center}\end{sloppypar}
\vspace{2mm}
\begin{sloppypar}
\noindent
R.W.~Clifft,
T.R.~Edgecock,
P.R.~Norton,
I.R.~Tomalin
\nopagebreak
\begin{center}
\parbox{15.5cm}{\sl\samepage
Particle Physics Dept., Rutherford Appleton Laboratory,
Chilton, Didcot, Oxon OX11 OQX, United Kingdom$^{10}$}
\end{center}\end{sloppypar}
\vspace{2mm}
\begin{sloppypar}
\noindent
\mbox{B.~Bloch-Devaux},$^{1}$
P.~Colas,
S.~Emery,
W.~Kozanecki,
E.~Lan\c{c}on,
\mbox{M.-C.~Lemaire},
E.~Locci,
P.~Perez,
J.~Rander,
\mbox{J.-F.~Renardy},
A.~Roussarie,
\mbox{J.-P.~Schuller},
J.~Schwindling,
A.~Trabelsi,$^{21}$
B.~Vallage
\nopagebreak
\begin{center}
\parbox{15.5cm}{\sl\samepage
CEA, DAPNIA/Service de Physique des Particules,
CE-Saclay, F-91191 Gif-sur-Yvette Cedex, France$^{17}$}
\end{center}\end{sloppypar}
\vspace{2mm}
\begin{sloppypar}
\noindent
N.~Konstantinidis,
A.M.~Litke,
G.~Taylor
\nopagebreak
\begin{center}
\parbox{15.5cm}{\sl\samepage
Institute for Particle Physics, University of California at
Santa Cruz, Santa Cruz, CA 95064, USA$^{13}$}
\end{center}\end{sloppypar}
\vspace{2mm}
\begin{sloppypar}
\noindent
C.N.~Booth,
S.~Cartwright,
F.~Combley,$^{4}$
M.~Lehto,
L.F.~Thompson
\nopagebreak
\begin{center}
\parbox{15.5cm}{\sl\samepage
Department of Physics, University of Sheffield, Sheffield S3 7RH,
United Kingdom$^{10}$}
\end{center}\end{sloppypar}
\vspace{2mm}
\begin{sloppypar}
\noindent
K.~Affholderbach,$^{28}$
A.~B\"ohrer,
S.~Brandt,
C.~Grupen,
A.~Ngac,
G.~Prange,
U.~Sieler
\nopagebreak
\begin{center}
\parbox{15.5cm}{\sl\samepage
Fachbereich Physik, Universit\"at Siegen, D-57068 Siegen,
 Germany$^{16}$}
\end{center}\end{sloppypar}
\vspace{2mm}
\begin{sloppypar}
\noindent
G.~Giannini
\nopagebreak
\begin{center}
\parbox{15.5cm}{\sl\samepage
Dipartimento di Fisica, Universit\`a di Trieste e INFN Sezione di Trieste,
I-34127 Trieste, Italy}
\end{center}\end{sloppypar}
\vspace{2mm}
\begin{sloppypar}
\noindent
J.~Rothberg
\nopagebreak
\begin{center}
\parbox{15.5cm}{\sl\samepage
Experimental Elementary Particle Physics, University of Washington, Seattle, 
WA 98195 U.S.A.}
\end{center}\end{sloppypar}
\vspace{2mm}
\begin{sloppypar}
\noindent
S.R.~Armstrong,
K.~Berkelman,
K.~Cranmer,
D.P.S.~Ferguson,
Y.~Gao,$^{20}$
S.~Gonz\'{a}lez,
O.J.~Hayes,
H.~Hu,
S.~Jin,
J.~Kile,
P.A.~McNamara III,
J.~Nielsen,
Y.B.~Pan,
\mbox{J.H.~von~Wimmersperg-Toeller},
W.~Wiedenmann,
J.~Wu,
Sau~Lan~Wu,
X.~Wu,
G.~Zobernig
\nopagebreak
\begin{center}
\parbox{15.5cm}{\sl\samepage
Department of Physics, University of Wisconsin, Madison, WI 53706,
USA$^{11}$}
\end{center}\end{sloppypar}
\vspace{2mm}
\begin{sloppypar}
\noindent
G.~Dissertori
\nopagebreak
\begin{center}
\parbox{15.5cm}{\sl\samepage
Institute for Particle Physics, ETH H\"onggerberg, HPK, 8093 Z\"urich, Switzerland.}
\end{center}\end{sloppypar}
}
\footnotetext[1]{Also at CERN, 1211 Geneva 23, Switzerland.}
\footnotetext[2]{Now at Universit\'e de Lausanne, 1015 Lausanne, Switzerland.}
\footnotetext[3]{Also at Dipartimento di Fisica di Catania and INFN Sezione di
 Catania, 95129 Catania, Italy.}
\footnotetext[4]{Deceased.}
\footnotetext[5]{Also Istituto di Cosmo-Geofisica del C.N.R., Torino,
Italy.}
\footnotetext[6]{Now at Institut f\"ur Experimentelle Kernphysik, Universit\"at Karlsruhe, 76128 Karlsruhe, Germany.}
\footnotetext[7]{Supported by CICYT, Spain.}
\footnotetext[8]{Supported by the National Science Foundation of China.}
\footnotetext[9]{Supported by the Danish Natural Science Research Council.}
\footnotetext[10]{Supported by the UK Particle Physics and Astronomy Research
Council.}
\footnotetext[11]{Supported by the US Department of Energy, grant
DE-FG0295-ER40896.}
\footnotetext[12]{Now at Departement de Physique Corpusculaire, Universit\'e de
Gen\`eve, 1211 Gen\`eve 4, Switzerland.}
\footnotetext[13]{Supported by the US Department of Energy,
grant DE-FG03-92ER40689.}
\footnotetext[14]{Also at Rutherford Appleton Laboratory, Chilton, Didcot, UK.}
\footnotetext[15]{Permanent address: Universitat de Barcelona, 08208 Barcelona,
Spain.}
\footnotetext[16]{Supported by the Bundesministerium f\"ur Bildung,
Wissenschaft, Forschung und Technologie, Germany.}
\footnotetext[17]{Supported by the Direction des Sciences de la
Mati\`ere, C.E.A.}
\footnotetext[18]{Supported by the Austrian Ministry for Science and Transport.}
\footnotetext[19]{Now at SAP AG, 69185 Walldorf, Germany.}
\footnotetext[20]{Also at Department of Physics, Tsinghua University, Beijing, The People's Republic of China.}
\footnotetext[21]{Now at D\'epartement de Physique, Facult\'e des Sciences de Tunis, 1060 Le Belv\'ed\`ere, Tunisia.}
\footnotetext[22]{Now at Groupe d' Astroparticules de Montpellier, Universit\'e de Montpellier II, 34095 Montpellier, France.}
\footnotetext[23]{Also at Dipartimento di Fisica e Tecnologie Relative, Universit\`a di Palermo, Palermo, Italy.}
\footnotetext[24]{Now at CERN, 1211 Geneva 23, Switzerland.}
\footnotetext[25]{Now at SLAC, Stanford, CA 94309, U.S.A.}
\footnotetext[26]{Now at INFN Sezione di Roma II, Dipartimento di Fisica, Universit\'a di Roma Tor Vergata, 00133 Roma, Italy.} 
\footnotetext[27]{Now at LBNL, Berkeley, CA 94720, U.S.A.}
\footnotetext[28]{Now at Skyguide, Swissair Navigation Services, Geneva, Switzerland.}
\footnotetext[29]{Now at Honeywell, Phoenix AZ, U.S.A.}
\setlength{\parskip}{\saveparskip}
\setlength{\textheight}{\savetextheight}
\setlength{\topmargin}{\savetopmargin}
\setlength{\textwidth}{\savetextwidth}
\setlength{\oddsidemargin}{\saveoddsidemargin}
\setlength{\topsep}{\savetopsep}
\normalsize
\newpage
\pagestyle{plain}
\setcounter{page}{1}

\section{Introduction}

Four orbitally-excited {\dsd} mesons, with angular 
momentum $L=1$, are expected to exist in addition to the pseudoscalar and 
vector mesons {\ds} and {\dsst}.
In the Heavy Quark Effective Theory \cite{HQET,HQET2}
the spin of the 
light quark couples with the orbital angular momentum to give 
$j=\frac{1}{2}$ or $j=\frac{3}{2}$. 
When coupled to the spin of the heavy quark, two doublets 
are obtained with the 
quantum numbers $J^P=0^+,1^+$ and $J^P=1^+,2^+$, respectively. 
These states are expected to decay mainly into D~K or D$^\ast$~K modes.

Only certain S- and D-wave decays are allowed by 
spin and parity conservation.  
The states of the $(j=\frac{1}{2})$ doublet decay via S-wave transitions
and are therefore expected to be broad ($\Gamma>100$ \Mev); 
they have not been observed yet. 
The $(j=\frac{3}{2})$ doublet states are much narrower since they can only
decay via D-wave modes.  
The two narrow states\footnote{The notation used follows that of the Particle Data Group 
\cite{pdg} for the {\dstst} mesons.} 
{\dsonepm} and {\dstwostpm} have been observed by
 ARGUS and CLEO
\cite{arg1,cleo1,cleo2}. At LEP, {\dsonepm} mesons have been 
observed by OPAL \cite{opal}.
The properties of the  four {\dsd} states are listed in \mbox{Table \ref{dsdprop}}. 
The actual physical particles could be superpositions 
of the individual states.

In this analysis the production rates of the two narrow states 
are measured in the decay modes\footnote{Throughout this letter 
the notation used for particles implies the charge conjugate modes as well.} 
{\dsda}, {\dsdb} and {\dsdc}. 
In {\Zcc} events, {\dsd} mesons
are produced in the fragmentation of primary c-quarks, whereas in \Zbb
events they can only be produced in decays of b~hadrons. To study 
these two
contributions separately, b- and c-quark-enriched event samples are 
also used.
  
\begin{table}[hhh]
\centerline{
\begin{tabular}{|l|c|c|c|c|c|c|c|}\hline
& & & & & & &\\[-0.9pc]
       & $J^P$    &  \Do~\Kp      & \Dp~\Ko   & \dsto~\Kp & \dstp~\Ko 
              & Mass [\Mevcc] & \mbox{Width [\Mev]}\\[0.1pc] \hline \hline
& & & & & & &\\[-0.9pc]
\dsonep  &       $1^+$  & \bigtimes & \bigtimes & D-wave & D-wave  
              & \mbox{$2535.3\pm 0.6$} \cite{pdg} & \mbox{$< 2.3$ (95\% CL)}  \cite{pdg} \\[0.1pc] \hline
& & & & & & &\\[-0.9pc]
\dszerostp &     $0^+$ & S-wave & S-wave & \bigtimes & \bigtimes 
            & -  & $>100$ \cite{Ddec} \\[0.1pc] \hline
& & & & & & &\\[-0.9pc]
\dsonestp  &    $1^+$ & \bigtimes & \bigtimes & S-wave & S-wave  
           & - & $> 100$ \cite{Ddec} \\[0.1pc] \hline
& & & & & & &\\[-0.9pc]
\dstwostp &      $2^+$ & D-wave & D-wave & D-wave & D-wave 
              & \mbox{$2573.5\pm 1.7$} \cite{pdg} & \mbox{$15^{+5}_{-4}$} \cite{pdg} \\[0.1pc] \hline
\end{tabular}}
\label{dsd_param}
\caption{\label{dsdprop} Masses and decay modes of {\dsd} mesons.} 
\end{table}

\section{The ALEPH detector} \label{detector}
The ALEPH detector and its performance 
are described in detail elsewhere~\cite{ALEPH-DETECTOR,ALEPH-PERFORMANCE}.
Only a brief
overview of the apparatus is given here. Surrounding the beam pipe, 
a high resolution vertex detector (VDET) consists of two layers of
double-sided silicon microstrip detectors,
positioned at average radii of 6.5~cm and 11.3~cm,
and covering respectively 85\% and 69\% of the solid angle.
The spatial resolution for the $r\phi$ and
$z$ projections (transverse to and along the beam axis, respectively) is
12~$\mu$m at normal incidence. The vertex detector is surrounded
by a drift chamber with eight coaxial wire layers with an outer
radius of 26~cm and by a time projection chamber
(TPC) that measures up to 21~three-dimensional points per track at radii
between 30~cm and 180~cm. These detectors are immersed
 in an axial magnetic field of 1.5~T provided by a superconducting 
solenoidal coil
and together measure the transverse momenta of
charged particles with a resolution
\mbox{$\sigma (p_{\mathrm T})/p_{\mathrm T} = 6 \times 10^{-4} \, p_{\mathrm T} \oplus 0.005$} 
($p_{\mathrm T}$ in \Gevc). The resolution of the
three-dimensional impact parameter for
tracks having information from all tracking
detectors and two VDET hits (a VDET ``hit'' being defined as a space point 
reconstructed from 
the $r\phi$ and $z$ coordinates) can be parametrized as
\mbox{$\sigma = 25\, \mu{\mathrm m} + 95\, \mu{\mathrm m}/p$} ($p$ in \Gevc ).
The TPC also provides up to 338~measurements of the specific ionization
of a charged particle.
The TPC is surrounded by a lead/proportional-chamber electromagnetic
calorimeter segmented into \mbox{$0.9^{\circ} \times 0.9^{\circ}$} projective towers
and read out in three sections in depth, with energy resolution
\mbox{$\sigma (E)/E = 0.18/\sqrt{E} + 0.009$} ($E$ in \Gev ). The iron return
yoke of the magnet is instrumented with streamer tubes to
form a hadron calorimeter, with a thickness of over 7
interaction lengths, and is surrounded by two layers of muon chambers.
An algorithm combines all these measurements to provide a determination
of the energy flow~\cite{ALEPH-PERFORMANCE} with an uncertainty
on the measurable total energy of
\mbox{$\sigma(E) = (0.6\sqrt{E/\mathrm{GeV}}+ 0.6)$~\Gev.}

\subsection{Event and track selection}

This analysis uses 4.1 million hadronic events 
recorded by the ALEPH detector at centre-of-mass energies
close to the {\Zo}~mass. The events are selected with the charged particle 
requirements described in~\cite{ALEPH-HADRONIC}.
The helix fit of the charged particle tracks used for the \dsd~reconstruction
must have a  $\chi^2$ per degree of freedom smaller than 5, and their polar
angle must satisfy $|\cos \theta| <$~0.95.
Finally the distance to the primary vertex in the
plane transverse to the beam axis   
has to be less than 2 cm and in the beam  direction less than 10 cm.

\subsection{Particle identification \label{dedx}}

Charged kaons are identified by means of the 
specific ionization loss {\it dE/dx} in the TPC.
The TPC provides two different measurements of the 
deposited energy, from the wire and the pad readout.
The wire readout is only used if at least 50 individual wire samples are
available. In this case the pad readout is ignored. If the wire information
is insufficient the pad readout is exploited. 
In both cases 
the particle identification is based on 
the {\it dE/dx} estimator $r_\pi$ ($r_{\rm K}$),
defined as the difference between the expected and the measured ionization loss 
expressed in terms of standard deviations for the pion (kaon) mass hypothesis.
A track is accepted as a kaon if its momentum is greater than {1.5~\Gevc},
$|r_{\rm K}|$ is less than 2.5 and it satisfies \mbox{$r_{\rm K}+r_\pi < 0$}.

Neutral kaons are reconstructed in the deay mode {\Kos~\arw~$\pi^+\pi^-$} 
as described in \cite{k0sel}.
For the identification of the {\Kos} their long lifetime ($c \tau = 2.7$cm)
is exploited by only accepting neutral vertices with a distance
of at least 1.5 cm to the primary vertex and a \mbox{$\chi^2 < 13$} of the vertex fit.
The reconstructed mass of the {\Kos} candidate has to be within {12~\Mevcc} of the 
nominal {\Kos} mass.  

Charged pions are selected by requiring $|r_\pi|$ to be smaller than 2.5.
Neutral pions are reconstructed 
 by a mass-constrained fit to the detected photons in the 
electromagnetic calorimeter \cite{ALEPH-PERFORMANCE}.
The momentum of the {\pio} candidate has to be greater than {1.5~\Gevc} and the $\chi^2$ of the mass fit less than 8.  

\subsection{Selection of c- and b-quark-enriched event samples}
\label{sec:cbsel}

To determine the production rates of {\dsd} in {\Zcc} and {\Zbb} events 
separately, c- and  
b-quark-enriched event samples have been selected. Due to the good spatial
resolution of the vertex detector, the probability (\Pevt) 
that all charged tracks originate from the primary vertex, determined with 
the algorithm described in \cite{qipbtag}, can be used to
separate long-lived b hadrons (low probability \Pevt ) 
from c and uds events.

An additional separation between c and b events is obtained by a cut on
the scaled energy $x_E$ (normalized to the beam energy) 
of the {\dsd} candidate, because less energy is
available for c quarks originating from a b-hadron decay than 
in direct {\Zcc} decays.

A {\Do} is reconstructed in all the decay modes considered. In order 
to further separate
light (uds) from heavy (c and b) quark events, a cut on the proper decay
time  of the {\Do} candidate is performed. 
The proper decay time is determined from the decay length $l$, calculated
as the
distance  of the {\Do} decay vertex from the primary vertex,
projected along the momentum 
vector of the {\Do}, and taking $t=M_{\mathrm D^0} \cdot l/p$. 

The cuts used to select the enriched samples 
are listed in \mbox{Table \ref{bccuts}}. About 4\% of all events are rejected
because they pass the cuts for the c- as well as the b-quark selection.

\begin{table}[h]
\centering
\begin{tabular}{|c|c|} \hline  
 & \\[-0.9pc]
 c-quark-enriched sample & b-quark-enriched sample\\[0.1pc] \hline \hline
 & \\[-0.9pc]
  0.1~ps $< t < $1.5~ps  & $t >$ 0.4~ps \\[0.1pc] \hline
 & \\[-0.9pc] 
  $-\log_{10}(${\Pevt}$) <$ 4     & $-\log_{10}(${\Pevt}$) >$ 1.5   \\[0.1pc]\hline
 & \\[-0.9pc]
  $ x_E > 0.4$  &  $0.25 < x_E < 0.65$ \\[0.1pc]\hline
\end{tabular}
\caption{\label{bccuts} Cuts for the selection of the c- and b-quark-enriched
event sample.}
\end{table}
\section{Reconstruction of {\boldmath \dsonep~and \dstwostp}}

\subsection{Decay channel {\boldmath \dsonep ~\arw~ \dstp \Ko}} 
\label{sec:dsd1}

\begin{figure}[tb]
\vspace{-0.2cm}
\centering
\epsfig{width=0.9\textwidth,file=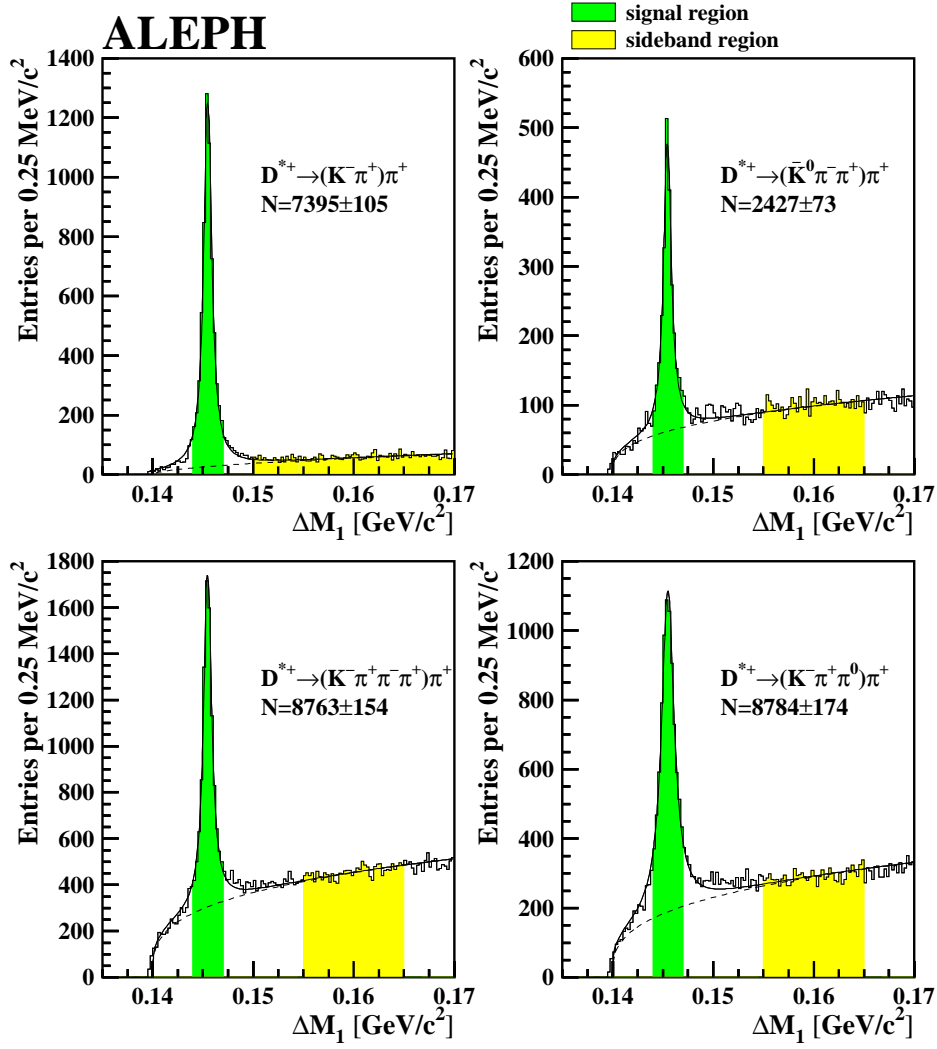}
\vspace{-0.7cm}
\caption{\label{dstplot} 
 Reconstructed  mass-difference  
 distribution of \mbox{$\Delta M_1$}. The quoted numbers of {\dstp} 
candidates are obtained from a fit to the distribution.}
\end{figure}

The {\dstp} mesons are reconstructed in the decay channel \mbox{\dstp ~\arw~ \Do \pip}
and {\Do} mesons in the four
decay channels {\docha}, {\dochb}, {\dochc} 
 and {\dochd}.
Because the combinatorial background is suppressed 
by the good resolution on the mass-difference
between the {\dstp} and {\Do} meson, the cuts
for the selection of {\Do} mesons are rather loose.
  
\begin{figure}[t]
\vspace{-1.0cm}
\centering
\epsfig{width=1.0\textwidth,file=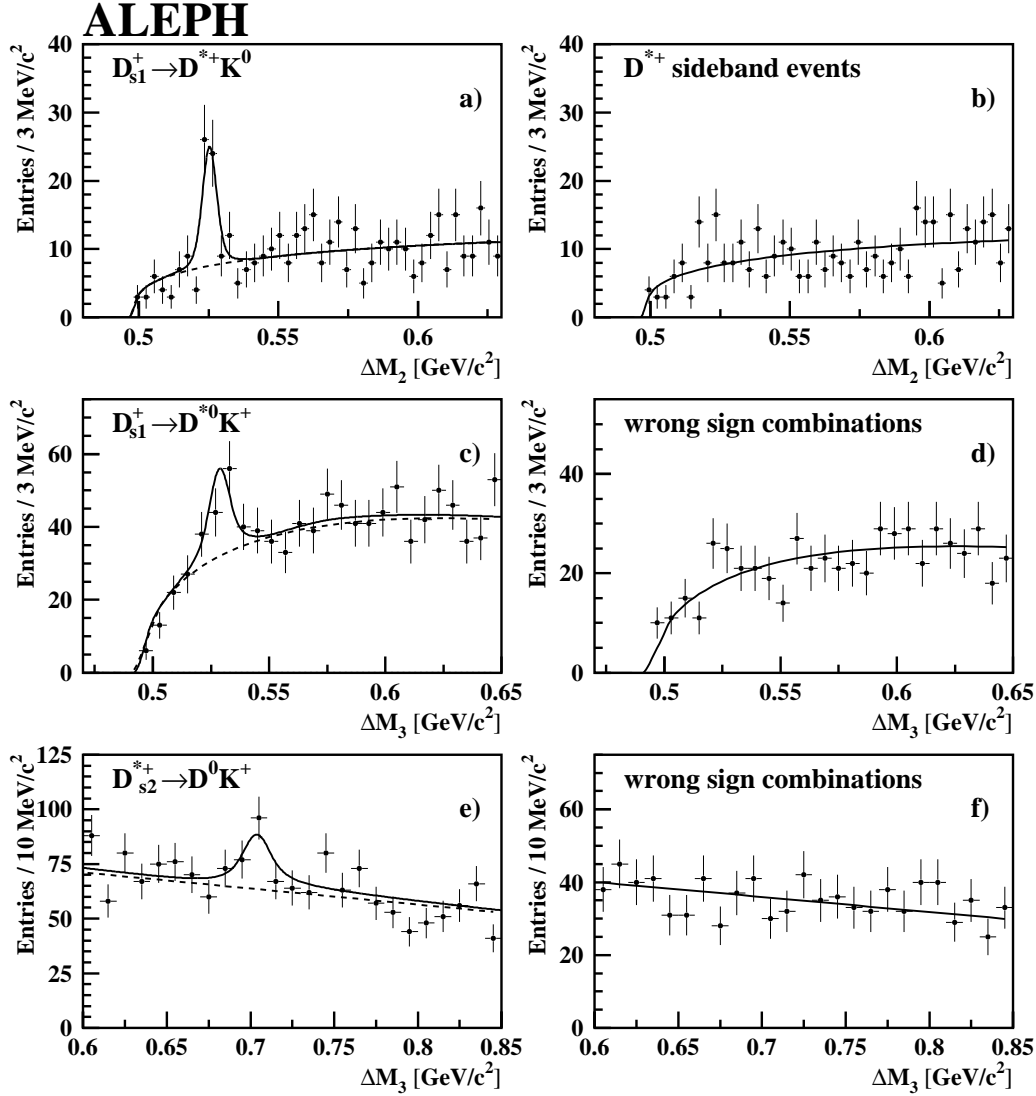}
\vspace{-0.7cm}
\caption{\label{dsdqq} 
 Reconstructed mass-difference  
 distribution for the decays \dsda (a and b), \dsdb (c and d) and \dsdc (e and f). 
 The full curve is the result of an unbinned log-likelihood fit and the 
dashed curve represents the combinatorial background.}
\end{figure}

To avoid double counting in the {\Do} decay channels with 
charged kaons and pions,  
the track with the smallest value of $r_{\rm K}$ is assumed to be the kaon track. 
A fit to a common decay vertex is performed for the
tracks of the selected charged particles and the $\chi^2$ has to be less than 20.
 The momentum of the 
{\Do} candidate has to be greater than {7~\Gevc}.
Additional cuts are listed in \mbox{Table \ref{docut}}.
\begin{table}[b]
\centering
\begin{tabular}{|l|c|c|c|c|} \hline
& \multicolumn{4}{|c|}{Decay channel} \\     
\multicolumn{1}{|c|}{variable} & \docha & \dochb & \dochc & \dochd\\ \hline\hline
$\Delta M_{{\rm D}^0}$ & $\pm$~20~\Mevcc &$\pm$~20~\Mevcc &$\pm$~15~\Mevcc &$\pm$~25~\Mevcc \\
\hline
$p_{\mathrm K}$ & $>$ 1.5 \Gevc & $>$ 3.0 \Gevc & $>$ 2.0 \Gevc & $>$ 2.0 \Gevc \\ 
\hline
$p_{\pi}$ & $>$ 0.7 \Gevc & $>$ 1.0 \Gevc & $>$ 0.4 \Gevc & $>$ 2.0 \Gevc  \\
\hline
$p_{\pi}$ (2nd) & --- &  $>$ 1.0 \Gevc & $>$ 1.5 \Gevc & ---  \\
\hline
$p_{\pi}$ (3rd) & --- & --- & $>$ 2.0 \Gevc & ---  \\
\hline
\end{tabular}
\vspace{0.0cm}
\caption{\label{docut} Cuts for the reconstruction of 
{\Do} mesons in the decay channels {\docha}, {\dochb}, {\dochc} and
{\dochd}.}
\end{table}
The selected {\Do} candidates are combined with all charged pions 
with a momentum less than {3~\Gevc}. 
For the {\dstp} candidates with a momentum greater than {8~\Gevc},
the mass difference \mbox{$\Delta M_1 =  m$(\Do,\pip) $- m$(\Do)} is calculated.    
The $\Delta M_1$ distributions for the four {\Do} decay channels 
are shown in Fig. \ref{dstplot}. 
The number of reconstructed {\dstp} mesons is determined by a fit to the measured 
$\Delta M_1$ distributions. The signal is parametrized as a 
Breit-Wigner and 
the background shape with the following function:\\
\begin{displaymath}
 \frac{dN}{d(\Delta M)} \propto (\Delta M-a)^b \cdot e^{-c(\Delta M-a)}.
\end{displaymath}
Each {\dstp} candidate inside the signal region ($\pm$1.5~\Mevcc) 
is combined with all {\Kos}  
candidates with a momentum greater than 2.5~\Gevc. 
The combination of the {\dstp}
 and the {\Kos} is required to have a scaled energy $x_E$ greater 
than 0.25. 
The mass difference 
\mbox{$\Delta M_2 = m$(\dstp,\Ko)$-m$(\dstp)$-m$(\Ko)$+m_{\mathrm{true}}$(\Ko)}
is calculated, where $m_{\mathrm{true}}$(\Ko) is the nominal {\Ko}  mass,
because this quantity has the best resolution.  
The resulting mass difference distribution $\Delta M_2$ is shown in
\mbox{Fig. \ref{dsdqq}a}. 
The distribution shows an excess of events in the signal
region around \mbox{525 \Mevcc }.  
The combinatorial background  (\mbox{Fig. \ref{dsdqq}b}) 
is determined with the help of {\dstp} 
candidates from sidebands above the {\dstp} mass window (Fig. \ref{dstplot}).
The number of signal events and the mass difference $\Delta M_2$ are 
determined by a simultaneous unbinned log-likelihood fit
to the signal and background distributions, with the decay width of the {\dsonep}
fixed to 1.5~\Mev.
The signal is fitted with a Breit-Wigner function, convoluted with 
the detector resolution. The detector resolutions for the reconstruction of
{\dsonep} mesons in the different {\Do} decay channels are determined
from simulated events. The resolution is approximately 1.7~\Mevcc for the 
{\Do} decay channels \docha, \dochb and \dochc and 2.6 \Mevcc for the
channel \dochd. 

The results of the fit to the $\Delta M_2$ distribution are listed in 
\mbox{Table \ref{dsd1res}}.  
The systematic errors on the mass differences are obtained by 
varying, in the fit, the assumed natural width of the {\dsonep} meson
and the detector resolution, as described in 
\mbox{Section \ref{systematics}}.

\subsection{Decay channels \mbox{\boldmath \dsonep ~\arw~ \dsto \Kp} and 
\mbox{\boldmath \dstwostp ~\arw~ \Do \Kp}}
\label{sec:dsd2}

\begin{figure}[tb]
\vspace{-1.0cm}
\centering
\epsfig{width=0.62\textwidth,file=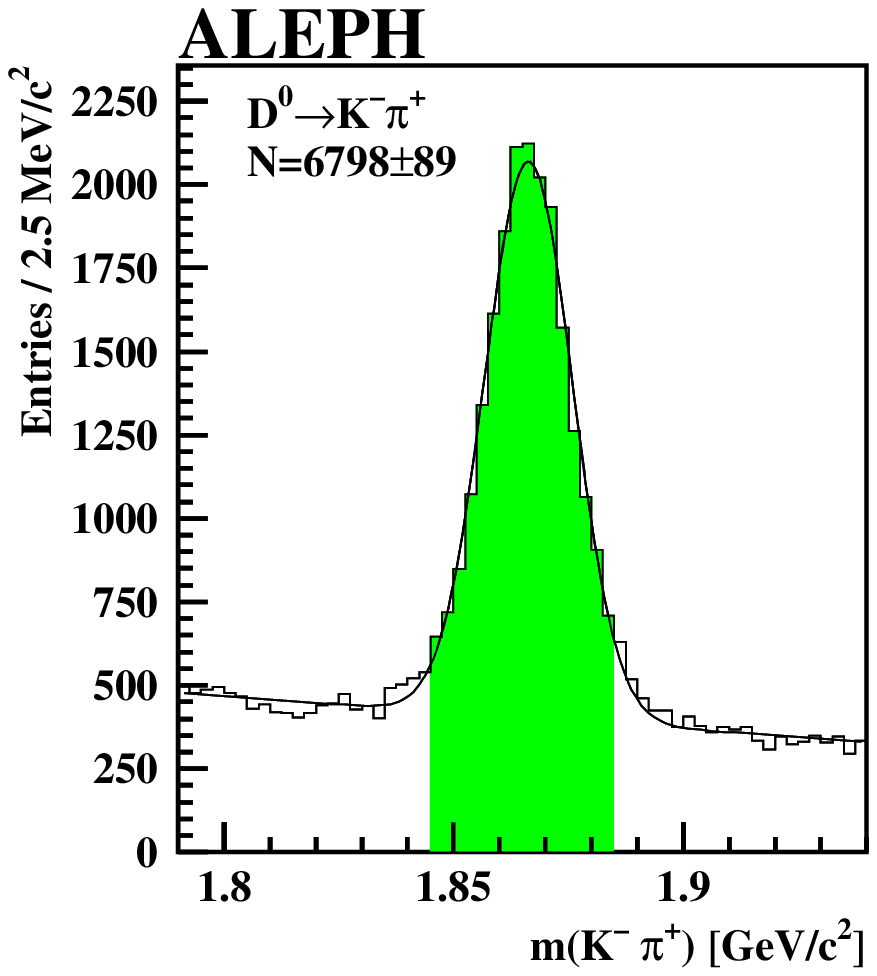}
\vspace{-0.5cm}
\caption{\label{D0mass} Invariant mass of the {\docha} candidates. The events in the 
shaded region are used for the 
reconstruction of the \mbox{\dsonep ~\arw~ \dsto \Kp} and \mbox{\dstwostp ~\arw~ \Do \Kp} decays.}
\end{figure}

The {\dsto} mesons decay into a {\Do} by emitting a {\pio}
or a photon. 
Because of the higher background in these channels the selection of {\Do}
candidates has to be  more restrictive than
for {\Do} candidates used in the reconstruction of {\dstp} mesons.
An acceptable signal-to-background ratio is only achieved in the decay channel {\docha} 
and therefore only this channel was used for the analysis.
The {\Do} candidates are rejected if the invariant mass of the {\Do} and 
any {\pip} in the event is within 2~\Mevcc of the {\dstp} mass window. 
The cuts on the momentum of the {\Do} candidate and on that of the decay 
particles are tightened to {8~\Gevc} and {3~\Gevc}, respectively. The {\Do}
candidate is also required to have a proper decay time greater than 0.2~ps.
The latter cut removes combinatorial background, peaked at small proper time.
The invariant mass distribution of the selected {\Do} candidates is shown 
in \mbox{Fig. \ref{D0mass}}.
The {\pio} or photon from the {\dsto} decay is not reconstructed in this
analysis. 
\begin{table}[tb]
\centerline{
\begin{tabular}{|c|c|c|} \hline
 & & \\[-0.9pc]
 Decay channel            & No. of events & $\Delta M$ [\Mevcc]  \\[0.1pc] \hline \hline
 & & \\[-0.9pc]
 \dsonep ~\arw~ \dstp \Ko  & $40.9 \pm 8.4_{\mathrm{stat}} {^{+3.2}_{-3.7}}_{\mathrm{syst}}$  
                          & $525.3 \pm 0.6_{\mathrm{stat}}\pm 0.1_{\mathrm{syst}}$ \\[0.1pc] \hline
 & & \\[-0.9pc]
 \dsonep \arw \dsto \Kp & $51.3 \pm 13.9_{\mathrm{stat}} {^{+3.8}_{-4.0}} _{\mathrm{syst}}$
                          & $528.7 \pm 1.9_{\mathrm{stat}}\pm 0.5_{\mathrm{syst}}$ \\[0.1pc] \hline
 & & \\[-0.9pc]
 \dstwostp \arw \Do \Kp & $63.6 \pm 21.9_{\mathrm{stat}} {^{+4.1}_{-9.5}} _{\mathrm{syst}}$
                        & $704 \pm 3_{\mathrm{stat}}\pm 1_{\mathrm{syst}}$ \\[0.1pc] \hline
\end{tabular}}
\caption{\label{dsd1res} Results of the fits to the flavour-independent sample.} 
\end{table}
Instead the combination of the {\Do}
with all charged kaons in the event with momentum greater than 3~{\Gevc} 
is taken and the mass difference \mbox{$\Delta M_3 = m$(\Do,\Kp)$- m$(\Do)}
is calculated.
Because of the small Q-value in the decay \mbox{\dsto ~\arw~ \Do \pio} of only 
{7~\Mev},
this procedure does not significantly change the resolution, 
which is determined from the simulation
to be \mbox{$(2.9 \pm 0.1)$ \Mevcc}. 
In the case where the {\dsto} emits a photon the
resolution is $(7.1 \pm 0.2)$ \Mevcc .
 
The {\dstwostp} meson is allowed to decay directly into the {\Do~\Kp} final
state and therefore the mass difference 
$\Delta M_3$ is also used to measure the decay \mbox{\dstwostp ~\arw~ \Do \Kp}.
The detector resolution in the reconstruction of this decay is 
$(3.1 \pm 0.1$)\Mevcc.  
The decay \mbox{\dstwostp ~\arw~ \dsto \Kp} is also allowed but suppressed 
due to the smaller phase space.
 
The measured mass-difference  distribution  
for {\Do~\Kp} combinations with a scaled energy greater than 0.25
is shown in \mbox{Figs. \ref{dsdqq}c-f}.
Signals for the {\dsdb} (\mbox{Fig. \ref{dsdqq}c}) 
and {\dsdc} decays (\mbox{Fig. \ref{dsdqq}e})
are visible, as expected, around 525~\Mevcc and {710~\Mevccns}, respectively.
The shape of the combinatorial background is determined with
the help of the wrong-sign charge combinations \Do~\Km (\mbox{Figs. \ref{dsdqq}d} and f).
The shape of the combinatorial background for the decay of the {\dsonep}
is parameterized with the function given in Section \ref{sec:dsd1}.
The combinatorial background for the decay of the {\dstwostp} is described
by a linear function of \mbox{$\Delta M_3$}.
The numbers of signal events are determined with a simultaneous
unbinned log-likelihood fit
to the signal and background distributions. The signal is
fitted with a Breit-Wigner function convoluted with the Gaussian  
detector resolution. In addition, reflections from the decays 
\mbox{D$_1^+ \rightarrow$ \dsto $\pi^+$} and \mbox{D$_2^{\ast +} \rightarrow $ \Do$ \pi^+$}
are taken into account in the fit to the {\dsdb} and {\dsdc} distributions. 
This contribution is 
small due to the broad distribution of the reflection events and is determined from simulated events to be 
4\% (6\%) of the  {\dsdb} ({\dsdc}) signal events.
The results of the fit to the $\Delta M_3$ distributions are listed in \mbox{Table \ref{dsd1res}}.
 The systematic uncertainties on the measured 
mass differences are obtained by varying, in the fit, the assumed 
decay width and the detector resolution, as described 
in Section \ref{systematics}.
 
From the measured mass differences, the mass of the {\dsonep} and 
{\dstwostp} can be calculated by adding the appropriate nominal mass
({\dstp}, {\dsto} or {\Do}) and calculating the weighted average 
in the case of the {\dsonep}. The resulting masses are  
\mbox{{\it m}(\dsonep)=$2535.3\pm 0.7$ \Mevcc} and 
\mbox{{\it m}(\dstwostp)=$2568.6\pm 3.2$ \Mevcc}.
The measured masses are in good agreement with the world averages 
given in \mbox{Table \ref{dsd_param}}. 

\subsection{Measurement of {\boldmath \dsd} decays in the c- and b-quark-enriched sample}

\begin{figure}[t]
\vspace{-1.0cm}
\centering
\epsfig{width=1.0\textwidth,file=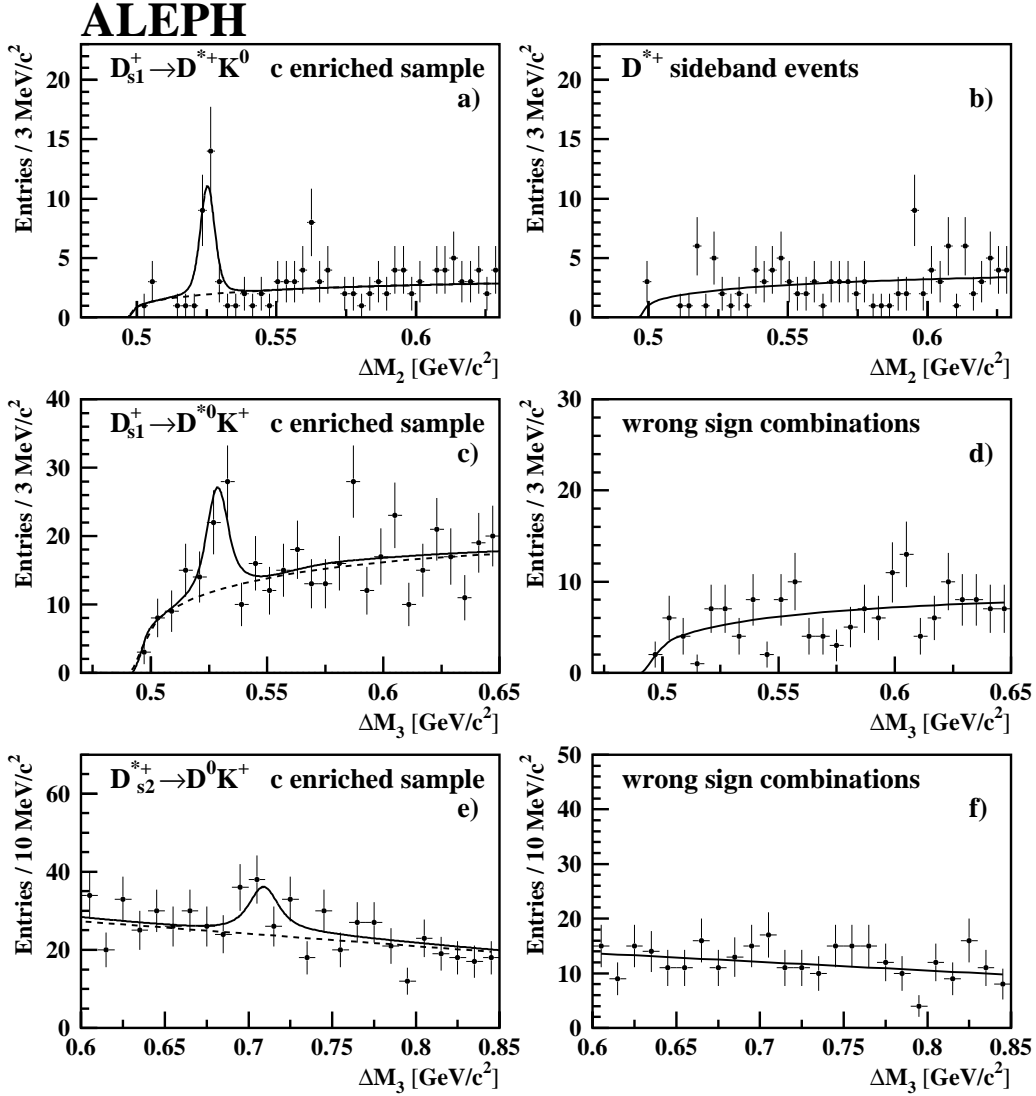}
\vspace{-0.7cm}
\caption{\label{dsdc} Reconstructed mass-difference  
 distribution for the c-quark-enriched event sample. 
 The full curve is the result of an unbinned log-likelihood fit and the 
dashed curve represents the combinatorial background.}
\end{figure}

\begin{figure}[t]
\vspace{-1.0cm}
\centering
\epsfig{width=1.0\textwidth,file=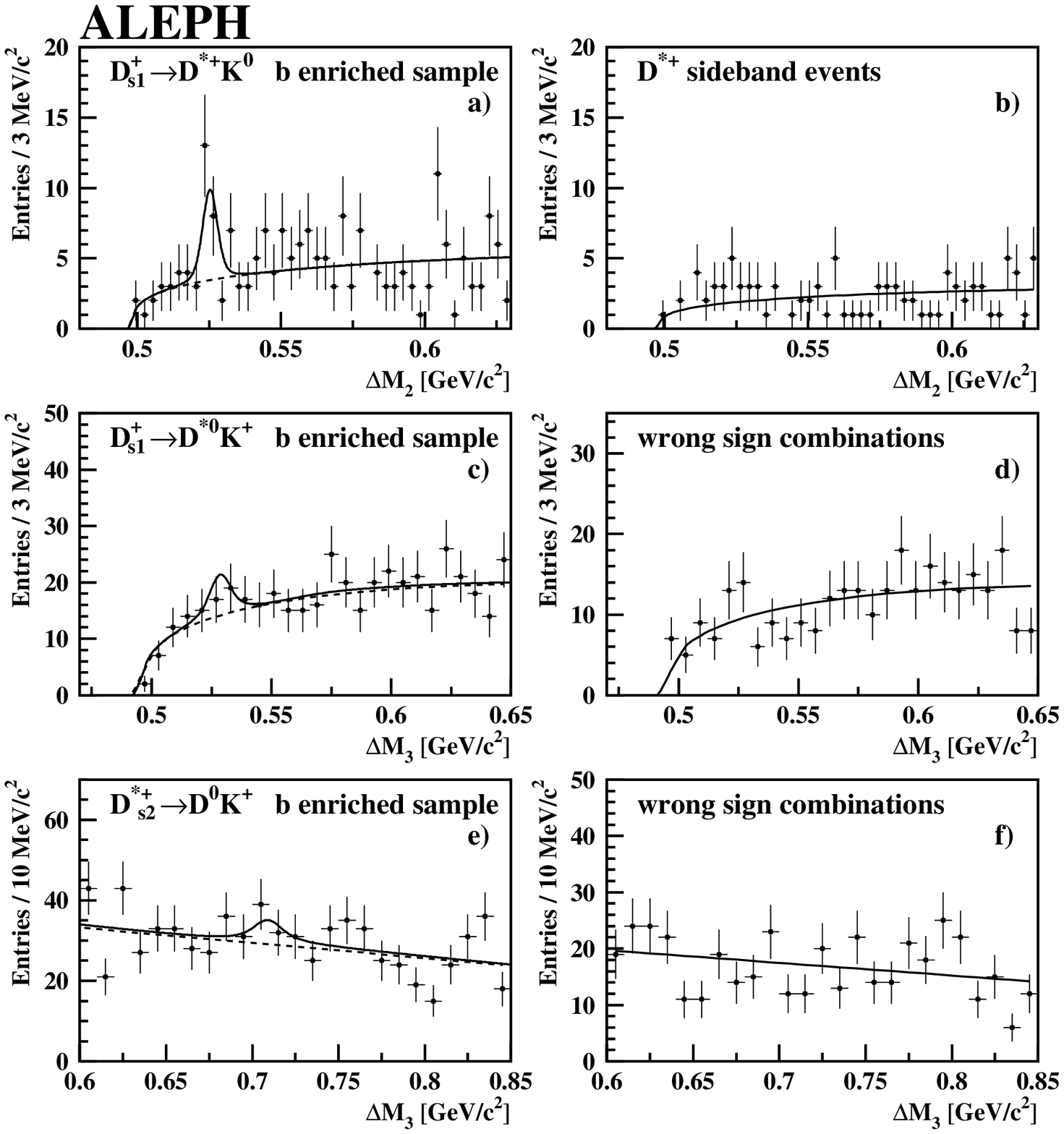}
\vspace{-0.7cm}
\caption{\label{dsdb} Reconstructed mass-difference 
 distribution for the b-quark-enriched event sample. 
 The full curve is the result of an unbinned log-likelihood fit to the data and the 
dashed curve represents the combinatorial background.}
\end{figure}

The measured mass-difference distributions for the 
c- and b-quark-enriched samples
are shown in Figs. \ref{dsdc} and \ref{dsdb}.
The {\dsone} and {\dstwost} mesons are reconstructed  
as described in the previous sections.
The numbers of signal events are determined with an unbinned 
log-likelihood fit to the data as described in the previous section, with the exception that 
the mass differences are fixed using the {\dsone} and {\dstwost} masses 
given in \mbox{Table \ref{dsdprop}}. 
The results are given in \mbox{Table \ref{bcresults}}.
\begin{table}[tb]
\centerline{
\begin{tabular}{|c|c|c|} \hline  
 & & \\[-0.9pc]
 Decay channel           & c-quark-enriched sample        & b-quark-enriched sample\\[0.1pc] \hline \hline
 & & \\[-0.9pc]
 \dsonep ~\arw~ \dstp \Ko & 
  $21.4 \pm 5.0_{\mathrm{stat}} {\pm~0.2} _{\mathrm{syst}}$ &
  $15.1 \pm 5.3_{\mathrm{stat}} {^{+0.8}_{-1.7}} _{\mathrm{syst}}$ \\[0.1pc] \hline
 & & \\[-0.9pc]
 \dsonep ~\arw~ \dsto \Kp  & 
  $32.2 \pm 9.3_{\mathrm{stat}} {\pm~1.3} _{\mathrm{syst}}$ &
  $15.4 \pm 7.8_{\mathrm{stat}} {\pm~1.8} _{\mathrm{syst}}$ \\[0.1pc] \hline
 & & \\[-0.9pc]
 \dstwostp ~\arw~ \Do \Kp  & 
  $30.7 \pm 14.1_{\mathrm{stat}} {^{+2.6}_{-5.2}} _{\mathrm{syst}}$ &
  $14.6 \pm 14.2_{\mathrm{stat}} {^{+2.6}_{-4.9}} _{\mathrm{syst}}$ \\[0.1pc] \hline
\end{tabular}}
\caption{\label{bcresults} Number of signal events in the c- and 
b-quark-enriched event sample.} 
\end{table}

\section{Reconstruction efficiencies}
\label{sec:eff}

\begin{table}[ht]
\begin{center}
\begin{tabular}{|l|c|c|c|} \hline
 & \\[-0.9pc]
{Decay channel} & {Mean efficiency}\\[0.1pc] \hline \hline
 & \\[-0.9pc]
{\dsda}~(\docha )          &  $(16.46 \pm 0.41) \%$ \\[0.1pc]\hline
 & \\[-0.9pc]
{\dsda}~(\dochb )          &  $(11.70 \pm 0.51) \%$ \\[0.1pc]\hline
 & \\[-0.9pc]
{\dsda}~(\dochc )          &  $(9.17 \pm 0.22) \%$ \\[0.1pc]\hline
 & \\[-0.9pc]
{\dsda}~(\dochd )          &  $(2.99 \pm 0.11) \%$ \\[0.1pc]\hline
 & \\[-0.9pc]
\dsdb                     &   $(11.00 \pm 0.14) \%$ \\[0.1pc]\hline
 & \\[-0.9pc]
\dsdc                     &  $(10.73 \pm 0.18) \%$ \\[0.1pc]\hline
\end{tabular}
\caption{\label{efficiencies} Mean reconstruction efficiencies 
of the decays {\dsda},
{\dsdb} and {\dsdc}.}
\end{center}
\end{table}

\begin{table}[h]
\begin{center}
\begin{tabular}{|l|c|c||c|c|} \hline
 & \multicolumn{2}{c||}{} & \multicolumn{2}{c|}{} \\[-0.9pc]
\raisebox{-1.5ex}{Decay channel} &
\multicolumn{2}{c||}{c-quark-enriched}&
\multicolumn{2}{c|}{b-quark-enriched} \\[0.1pc] \cline{2-5}
       & c ~\arw~ \dsd & b ~\arw~ \dsd 
       & c ~\arw~ \dsd & b ~\arw~ \dsd \\[0.1pc] \hline \hline
 & & & & \\[-0.9pc]
\dsda &  $(5.87 \pm 0.30) \%$ & 
         $(0.88 \pm 0.14) \%$ &
         $(0.34 \pm 0.04) \%$ & 
         $(4.08 \pm 0.20) \%$ \\[0.1pc]\hline
 & & & & \\[-0.9pc]
\dsdb &  $(9.14 \pm 0.22) \%$ & 
         $(1.40 \pm 0.06) \%$ &
         $(0.70 \pm 0.06) \%$ &
         $(7.81 \pm 0.15) \%$ \\[0.1pc]\hline
 & & & & \\[-0.9pc]
\dsdc &  $(9.99 \pm 0.18) \%$ &
         $(1.36 \pm 0.08) \%$  &
         $(0.59 \pm 0.05) \%$ & 
         $(7.34 \pm 0.17) \%$ \\[0.1pc]\hline
\end{tabular}
\caption{\label{cefficiencies} 
Reconstruction efficiencies for the decays {\dsda}, {\dsdb} and 
{\dsdc} in the c- and b-quark-enriched event samples.
For the decay {\dsda} the reconstruction efficiencies
are branching-ratio-weighted sums of the four different {\Do} decay 
modes.
The low efficiency for \mbox{b ~\arw~ \dsd} (\mbox{c ~\arw~ \dsd}) in the c-quark 
 (b-quark) enriched sample reflects the contamination of approximately 
\mbox{$10-15$\%} of the other flavour.}
\end{center}
\end{table}

To determine the production rates in hadronic events, the 
reconstruction efficiencies are taken from the simulation.
The efficiencies are calculated separately for {\dsd} mesons produced
in charm fragmentation, b-hadron decays and from gluon splitting.

In the flavour-independent selection the reconstruction efficiencies
for {\dsd} mesons produced in charm fragmentation and in b-hadron decays
are approximately the same.
The efficiency for {\dsd} mesons from 
gluon splitting is lower due to their softer energy spectrum. 
The mean efficiencies are
calculated from the flavour-specific efficiencies  
under the assumption that {\dsd} mesons are produced in 
equal amounts in charm fragmentation and in b-hadron decays. 
The fraction of {\dsd} mesons produced in gluon splitting is 
taken from the simulation, with the rate of gluons splitting into
c$\overline{{\rm c}}$ pairs reweighted to the measured value
of $f$(\gcc)=0.0296~$\pm$~0.0038 \cite{LEPEW2}, and found to be about 6~\%. 
The average efficiency is calculated as
\mbox{$\epsilon_{\mathrm m}=0.47 \cdot \epsilon_{\mathrm c} + 0.47 \cdot \epsilon_{\mathrm b}
+ 0.06 \cdot \epsilon_{\mathrm g}$}. The calculated mean efficiencies are given in
\mbox{Table \ref{efficiencies}}.

The reconstruction efficiencies in the c- and the b-quark-enriched 
event samples
for the different {\dsd} decays are listed in \mbox{Table \ref{cefficiencies}}.
 
\section{Production rates}
The production rates are calculated from the observed number of events,
the total number of hadronic \Zo~decays 
\mbox{($N_{\rm evt}$=4$\,$151$\,$890),} 
the known branching ratios and the reconstruction efficiencies
\begin{eqnarray}
{\it f}(\mbox{\Zo~\arw~\dsonepm}) \cdot \mbox{\Br(\dsda)} & = &
\frac{2 \cdot n_1}{N_{\rm evt} \cdot \epsilon_{1} \cdot
\mbox{\Br(\dstp~\arw~\Do \pip)} \cdot \mbox{\Br(\Kos~\arw~\pip \pim)}} \;, \\
{\it f}(\mbox{\Zo~\arw~\dsonepm}) \cdot \mbox{\Br(\dsdb)} & = & 
\frac{n_2}{N_{\rm evt} \cdot \epsilon_{2} \cdot
\mbox{\Br(\docha)}}  \;,\\
{\it f}(\mbox{\Zo~\arw~\dstwostpm}) \cdot \mbox{\Br(\dsdc)} & = & 
\frac{n_3}{N_{\rm evt} \cdot \epsilon_{3} \cdot
\mbox{\Br(\docha)}}\;,
\end{eqnarray}
where $n_1$, $n_2$ and $n_3$ denote the number of reconstructed 
events in the three
considered {\dsd} channels and $\epsilon_{1}$, $\epsilon_{2}$ and 
$\epsilon_{3}$
the corresponding branching-ratio-weighted sums of the 
reconstruction  efficiencies.
The branching ratios of the {\dstp}, {\Kos}, {\Do}, and those of all
subsequent decays, are taken from \cite{pdg}.
   
In the case of the c-  and b-quark-enriched samples, the rate is calculated per
event hemisphere and not per \Zo-decay. In this case the number of hadronic 
events are multiplied by the partial width 
 R$_\mathrm{c}$ and 
R$_\mathrm{b}$,
for which the Standard Model values \mbox{R$_\mathrm{c}=17.2\%$} and  
\mbox{R$_\mathrm{b}=21.6\%$} are used.
The average number of {\dsd} expected from gluon splitting ($1-2$\%) is 
subtracted from the number of events observed. 
Since the enriched samples are not pure, but have a contamination of 
roughly 10\% from other quark flavours, the production rates are 
determined simultaneously for the c- and b-quark-enriched samples.

\subsection{Systematic errors} 
\label{systematics}

\begin{table}[p]
\centering
\begin{tabular}{|l|c|c|c|} \hline 
\multicolumn{4}{|c|}{} \\[-0.9pc] 
\multicolumn{4}{|c|}{Systematic errors in the decay \dsda} \\[0.1pc] \hline \hline
Error source&
 flavour independent & c-enriched & b-enriched \\ \hline
Mass, decay width and resolution &
  $+8~-9$ & $ \pm~1$ &  $+5~-11$ \\ \hline
$x_E(\mathrm{c})$, $x_E(\mathrm{b})$  variation & 
  $\pm~1$ &  $\pm~1$ &  $\pm~1$ \\ \hline
Kaon  {\it dE/dx} selection & 
  $\pm~3$ & $\pm~3$ & $\pm~3$   \\ \hline
\dsd \ms source (gluon, c and b) &  
 $\pm~3$ &  $\pm~1$ &  $\pm~1$ \\ \hline
Monte Carlo statistics &
 $\pm~3$  & $\pm~3$  & $\pm~4$  \\ \hline 
{\Do} branching ratios &
 $\pm~5$ &  $\pm~5$ &  $\pm~5$ \\
 \hline \hline
Total &
 $+11~-12$ &  $\pm~7$ &  $+9~-13$ \\ \hline
\multicolumn{4}{c}{} \\[-0.5pc] \hline 
\multicolumn{4}{|c|}{} \\[-0.9pc] 
\multicolumn{4}{|c|}{Systematic errors in the decay {\dsdb}} \\[0.1pc] \hline \hline
Error source&
 flavour independent & c-enriched & b-enriched \\ \hline
Mass, decay width and resolution  & $\pm~8$ &  $\pm~4$ &  $\pm~12$ \\ \hline
$x_E(\mathrm{c})$, $x_E(\mathrm{b})$  variation & 
 $\pm~1$ &  $\pm~1$ &  $\pm~2$ \\ \hline
Kaon  {\it dE/dx} selection & 
 $\pm~3$ &  $\pm~3$ &  $\pm~3$ \\ \hline
D$^{\ast\ast}$ reflections & 
 $\pm~2$ &  $\pm~4$ &  $+3~-2$ \\ \hline
\dsd \ms source (gluon, c and b) &  
 $\pm~3$ &  $\pm~1$ &  $\pm~1$ \\ \hline
Monte Calo statistics &
 $\pm~1$ &  $\pm~1$ &  $\pm~1$ \\ \hline 
{\Do} branching ratio &
 $\pm~2$ &  $\pm~2$ &  $\pm~2$ \\
 \hline \hline
Total &
 $\pm~10$ &  $\pm~7$ &  $\pm~13$ \\ \hline
\multicolumn{4}{c}{} \\[-0.5pc] \hline 
\multicolumn{4}{|c|}{} \\[-0.9pc] 
\multicolumn{4}{|c|}{Systematic errors in the decay {\dsdc}} \\[0.1pc] \hline \hline
Error source&
 flavour independent & c-enriched & b-enriched \\ \hline
Mass, decay width and resolution  &
 $+7~-15$ &  $+9~-17$ &  $+18~-34$ \\ \hline
$x_E(\mathrm{c})$, $x_E(\mathrm{b})$  variation & 
 $\pm~1$ &  $\pm~1$ &  $\pm~2$ \\ \hline
Kaon  {\it dE/dx} selection & 
 $\pm~3$ &  $\pm~3$ &  $\pm~3$ \\ \hline
D$^{\ast\ast}$ reflections & 
 $\pm~3$ &  $\pm~3$ &  $\pm~14$ \\ \hline
\dsd \ms source (gluon, c and b) &  
 $\pm~3$ &  $\pm~1$ &  $\pm~1$ \\ \hline
Monte Carlo statistics &
 $\pm~1$ &  $\pm~1$ &  $\pm~1$ \\ \hline 
{\Do} branching ratio &
  $\pm~2$ &  $\pm~2$ &  $\pm~2$ \\ \hline 
\hline
Total &
 $+9~-16$ &  $+10~-18$ &  $+23~-37$ \\ \hline
\end{tabular}
\caption{\label{systot1} Systematic error contributions in \%.}
\end{table}

The main contribution to the systematic error originates from the 
uncertainty on the parameters held fixed in the fit to the mass 
distributions of the {\dsonep}
and {\dstwostp} candidates. 
In all fits the natural decay widths 
(\mbox{$\Gamma$(\dsonepm) = 1.5~\Mev} and 
\mbox{$\Gamma$(\dstwostpm) = 15~\Mev}) and the 
detector resolution are fixed. In order to determine the systematic error,
the decay width of the {\dsonepm} ({\dstwostpm}) is varied between
1 and 2~\Mev (10 and 20~\Mev). An uncertainty of $\pm$10\%  
is assumed for the detector resolution determined from the simulation.
 In the case of the c- and b-quark-enriched 
samples, the central values of the {\dsonep} and {\dstwostp} masses 
are fixed to the world averages \cite{pdg} and are varied by 
$\pm$0.6~\Mevcc for the {\dsonep} and by $\pm$2~\Mevcc for the 
{\dstwostp}.
Each parameter is varied independently and the change in the number
of signal events is taken as the systematic error from that particular 
source. The total systematic error listed in \mbox{Tables \ref{dsd1res}} and 
\ref{bcresults} is given by the quadratic sum of all the above contributions.

The rate of gluons splitting to c$\overline{{\rm c}}$ pairs is varied within 
the experimental uncertainties. In addition, the selection efficiency for {\dsd} 
mesons originating from this source is varied by 50\% to account for the 
modeling of the {\gcc} process. 
An error of 
20\% is assumed on the equality of the \dsd \ms production in c and 
b events, which is used in the calculation of the average efficiency 
(see Section \ref{sec:eff}). The resulting error on the production rate 
is 1\%.

The fraction of reflection events from D$^{\ast\ast}$ decays is taken 
from the simulation.
The obtained ratio between reflection and signal events is varied by
50\% and the 
corresponding change in the number of signal events taken as the systematic error.
 
In the simulation the parameters of the Peterson fragmentation 
function are reweighted to 
reproduce the measured values of the mean scaled energy $x_E$  
of c and b hadrons: \mbox{$x_E(\mathrm{c})=0.484 \pm 0.008$} and  
\mbox{$x_E(\mathrm{b})=0.702 \pm 0.008$ \cite{LEPEW}}, respectively.
The reweighted samples are used to determine the reconstruction 
efficiencies and the errors 
on the fragmentation parameters are propagated as well.

The  {\it dE/dx} selection efficiency determined from the simulation is tested 
by reconstructing {\dstp ~\arw~ (\Km \pip)\pip} decays without using the  {\it dE/dx}
 information of the \Km.
By applying subsequently the  {\it dE/dx} cut used in the analysis, 
the efficiency of 
this cut can be determined in data and the simulation.  
The difference between data and the simulation
is \mbox{($-3.8 \pm 2.9$)\%}.
The reconstruction efficiencies of the appropriate decay channels are corrected 
for the observed offset and the uncertainty is taken into account 
as a systematic error.

Finally the errors of the reconstruction efficiency due to the finite 
Monte Carlo statistics and the uncertainty on the used branching 
ratios are taken into account.

The total systematic error is calculated by adding the different
contributions in quadrature. The list of the systematic errors is given in
\mbox{Table \ref{systot1}}. 
The systematic errors of the two {\dsone}-modes have been assumed to be 100\%
correlated.

\subsection{Results} \label{sec:results}
From the measured numbers of produced {\dsonepm} and {\dstwostpm} mesons
and the reconstruction efficiencies, the production rates in the decay
channels {\dsda}, {\dsdb} and {\dsdc} 
in hadronic events are\\[-0.3cm]
\begin{center}
{\it f}$\,$(\Zo~\arw~\dsonepm) $\cdot$ \Br(\dsonep ~\arw~ \dstp \Ko)
 $=(0.22 \pm 0.05_{\mathrm{stat}} {\pm~0.03}_{\mathrm{syst}}) \%$ \\[0.2cm]
{\it f}$\,$(\Zo~\arw~\dsonepm) $\cdot$ \Br(\dsonep ~\arw~ \dsto \Kp)
 $=(0.29 \pm 0.08_{\mathrm{stat}} {\pm~0.03}_{\mathrm{syst}}) \%$ \\[0.2cm]
{\it f}$\,$(\Zo~\arw~\dstwostpm) $\cdot$ \Br(\dstwostp ~\arw~ \Do \Kp)
 $=(0.37 \pm 0.13_{\mathrm{stat}} {^{+0.03}_{-0.06}}_{\mathrm{syst}}) \%$ .\\
\end{center}
%
The production rates in the c-quark-enriched sample are\\[-0.3cm]
\begin{center}
{\it f}$\,$(c~\arw~\dsonepm) $\cdot$ Br(\dsonep~\arw~\dstp \Ko)
 $=(0.35 \pm 0.10_{\mathrm{stat}} {\pm~0.03}_{\mathrm{syst}}) \%$ \\[0.2cm]
{\it f}$\,$(c~\arw~\dsonepm) $\cdot$ Br(\dsonep~\arw~\dsto \Kp)
 $=(0.59 \pm 0.19_{\mathrm{stat}} {\pm~0.04}_{\mathrm{syst}}) \%$\\[0.2cm]
{\it f}$\,$(c~\arw~\dstwostpm) $\cdot$ Br(\dstwostp~\arw~\Do \Kp)
 $=(0.51\pm 0.27_{\mathrm{stat}}  {^{+0.05}_{-0.09}}_{\mathrm{syst}}) \%$ .\\\
\end{center}
The production rates in the b-quark-enriched sample are\\[-0.3cm]
\begin{center}
{\it f}$\,$(b~\arw~\dsonepm) $\cdot$ Br(\dsonep~\arw~\dstp \Ko)
 $=(0.30 \pm 0.12_{\mathrm{stat}} {^{+0.03}_{-0.04}}_{\mathrm{syst}}) \%$\\[0.2cm]
{\it f}$\,$(b~\arw~\dsonepm) $\cdot$ Br(\dsonep~\arw~\dsto \Kp)
 $=(0.24 \pm 0.15_{\mathrm{stat}} {\pm~0.03}_{\mathrm{syst}}) \%$\\[0.2cm]
{\it f}$\,$(b~\arw~\dstwostpm) $\cdot$ Br(\dstwostp~\arw~\Do \Kp)
  $=(0.25 \pm 0.29_{\mathrm{stat}} {^{+0.06}_{-0.09}}_{\mathrm{syst}}) \%$ .\\
\end{center}
The production rate of {\dstwostpm} mesons in b-decays has a significance of less than
one sigma, therefore 
a $95\%$ confidence limit is calculated by renormalizing the probability 
for the allowed region of the production rate
and taking into account the measured central value:\\[-0.5cm]
\begin{center}
{\it f}$\,$(b~\arw~\dstwostpm) $\cdot$ Br(\dstwostp~\arw~\Do \Kp) $<$
$0.77  \% \; $at $95$\% CL .\\
\end{center}
Assuming that the decay width of the {\dsonep} is saturated by the two
measured decays and
that the branching ratio Br(\dstwostp ~\arw~ \Do \Kp ) is $45\%$, which follows from the
assumption that \mbox{Br(\dstwostp ~\arw~ D K)=90\%} with  10\% 
branching ratio into \mbox{D$^\ast$ K}  \cite{Ddec}, 
the total production rates in hadronic events are found to be\\[-0.5cm]
\begin{center}
{\it f}$\,$(\Zo~\arw~\dsonepm)
 $=(0.52 \pm 0.09_{\mathrm{stat}} {\pm~0.06}_{\mathrm{syst}}) \%$\\[0.2cm]
{\it f}$\,$(\Zo~\arw~\dstwostpm)
 $=(0.83 \pm 0.29_{\mathrm{stat}} {^{+0.07}_{-0.13}}_{\mathrm{syst}}) \%$ .\\
\end{center}
The production rates from c-quarks (c ~\arw~ \dsd) are\\[-0.3cm] 
\begin{center}
{\it f}$\,$(c~\arw~\dsonepm)
 $=(0.94\pm 0.22_{\mathrm{stat}} {\pm~0.07}_{\mathrm{syst}}) \%$\\[0.2cm]
{\it f}(c~\arw~\dstwostpm)
 $=(1.14\pm 0.59_{\mathrm{stat}} {^{+0.11}_{-0.20}}_{\mathrm{syst}}) \%$ .\\
\end{center}
The production rates from b-quarks  (b ~\arw~ \dsd) are\\[-0.3cm]
\begin{center} 
{\it f}$\,$(b~\arw~\dsonepm)
 $=(0.55 \pm 0.19_{\mathrm{stat}} {^{+0.06}_{-0.07}}_{\mathrm{syst}}) \%$\\[0.2cm]
{\it f}$\,$(b~\arw~\dstwostpm)
  $=(0.57 \pm 0.63_{\mathrm{stat}} {^{+0.13}_{-0.20}}_{\mathrm{syst}}) \%$ .\\
\end{center}
The production rate of {\dstwostpm} mesons in b-decays has a significance of less than
one sigma and the corresponding upper limit is\\[-0.5cm]
\begin{center}
{\it f}$\,$(b~\arw~\dstwostpm) $<$
$1.70  \% \; $at $95$\% CL .
\end{center}

\section{Conclusions}

The production rates of the {\dsd} 
mesons {\dsonep} and {\dstwostp} have been
measured by reconstructing the decays {\dsda}, {\dsdb} and {\dsdc}.
 
In the two measured {\dsonep} decay modes an 
enhancement of the \mbox{\dsto \Kp} final state is expected due to 
the higher Q-value of the decay, since isospin invariance requires 
the matrix elements of the two decays to be the same. 
The expected ratio of the two branching ratios is given by \cite{Ddec}\\
\begin{displaymath}
\frac{\mbox{Br(\dsonep~\arw~\dsto \Kp)}}
       {\mbox{Br(\dsonep~\arw~\dstp \Ko)}} =
 \left( \frac{\mbox{q(\Kp)}}{\mbox{q(\Ko)}} \right) ^{2L+1} \approx 1.77\, ,
\end{displaymath}
where q(\Kp) (q(\Ko)) is the momentum of the \Kp (\Ko) in the rest frame of the
{\dsonep} and \mbox{$L$=2}  for a pure D-wave decay. 
 
The measured ratio of the two production rates is
\begin{displaymath}
\frac{\mbox{Br(\dsonep ~\arw~ \dsto \Kp)}}
      {\mbox{Br(\dsonep ~\arw~ \dstp \Ko)}}
 = 1.32 \pm 0.47_{\mathrm{stat}} \pm 0.23_{\mathrm{syst}} .
\end{displaymath}
The smaller value compared with the expectation could be due to an admixture of a S-wave 
decay from one of the broad 
{\dsd} states (see \mbox{Table \ref{dsdprop}}). 
However, no significant discrepancy with the expected production rate ratio for a pure
D-wave decay is observed.

\subsection*{Acknowledgements}

It is a pleasure to thank our colleagues in the accelerator divisions of CERN
for the excellent performance of LEP.
Thanks are also due to the technical
personnel of the collaborating institutions for their support in constructing
and maintaining the ALEPH experiment. Those of us not from member
states wish to thank CERN for its hospitality.

%
\end{document}